\documentclass[%
 reprint,
superscriptaddress,
 amsmath,amssymb,
 aps,
 prd,
]{revtex4-1}

\usepackage{graphicx}% Include figure files
\usepackage{dcolumn}% Align table columns on decimal point
\usepackage{bm}% bold math
\usepackage{hyperref}
\usepackage{enumerate}
\usepackage{amsmath,amstext}

\hypersetup{
    pdfnewwindow=true,      % links in new window
    colorlinks=true,       % false: boxed links; true: colored links
    linkcolor=blue,          % color of internal links
    citecolor=blue,        % color of links to bibliography
    filecolor=blue,      % color of file links
    urlcolor=blue           % color of external links
}
\providecommand{\aap}[0]{Astron. Astrophys. }

\providecommand{\astropartphys}[0]{Astropart. Phys. }
\providecommand{\apj}[0]{Astrophys. J.~}
\providecommand{\apjl}[0]{Astrophys. J. Lett. }
\providecommand{\apjs}[0]{Astrophys. J. Supp. Ser. }

\providecommand{\cqg}[0]{CQG~} % Classical and Quantum Gravity

\providecommand{\lrr}[0]{Living Rev. Relativ.} % Living Reviews of Relativity
\providecommand{\mnras}[0]{Mon. Not. Roy. Astron. Soc. }

\providecommand{\nature}[0]{Nature }
\providecommand{\prd}{Phys. Rev. D. }
\providecommand{\prl}[0]{Phys. Rev. Lett. }
\begin{document}

\title{Gravitational-Wave Localization Alone Probes AGN Origin of \\ Stellar-Mass Black Hole Mergers}
%\title{Gravitational-Wave Localization Alone Can Reveal Active-Galactic-Nucleus Origin of Stellar-Mass Binary Black Hole Mergers}

\author{I. Bartos}
\email{ibartos@phys.columbia.edu}
\affiliation{Columbia Astrophysics Laboratory, Columbia University, New York, NY 10027, USA}
\author{Z. Haiman}%
\affiliation{Department of Astronomy, Columbia University, New York, NY 10027, USA}
\affiliation{Department of Physics, New York University, New York, NY 10003, USA}
\author{Z. Marka}
\affiliation{Columbia Astrophysics Laboratory, Columbia University, New York, NY 10027, USA}
\author{B.D. Metzger}
\affiliation{Columbia Astrophysics Laboratory, Columbia University, New York, NY 10027, USA}
\author{N.C. Stone}
\email{Einstein Fellow}
\affiliation{Columbia Astrophysics Laboratory, Columbia University, New York, NY 10027, USA}
\author{S. Marka}
\affiliation{Columbia Astrophysics Laboratory, Columbia University, New York, NY 10027, USA}

\begin{abstract}
Stellar-mass binary black hole mergers are poised to represent the majority of gravitational-wave (GW) observations by Advanced LIGO and Virgo. Probing their origin will be difficult due to the expected lack of electromagnetic emission and limited localization accuracy. Associations with rare host galaxy types -- such as active galactic nuclei (AGN) -- can nevertheless be identified statistically through spatial correlation. %Hosts' spatial correlation with observed GWs from binary black hole (BBH) mergers can be used to determine their contribution to the overall GW detection rate.
We show that (i) fractional contributions $f_{\rm agn}=50-100\%$  from AGN hosts to the total BBH merger rate can be statistically established with 70-300 detected events (expected in 0.5-2 years of observation with Advanced LIGO-Virgo at design sensitivity and current rate estimates); (ii) fractional contributions as low as $f_{\rm agn}=25\%$ can be tested with 1000 events ($\sim$ 5\,years of observation); (iii) the $\sim5\%$ best localized GWs drive these constraints.
%ZH:  last sentence may need to be modified
%We describe the analysis method and present results for AGN. Our method can help distinguish other binary formation channels with rare host populations.
The presented method and results are generally applicable to binary formation channels with rare host populations.
\end{abstract}


% 04.30.Db	Gravitational Wave generation and sources
% 04.30.-w	Gravitational waves
% 04.30.Tv	Gravitational-wave astrophysics
% 04.80.Nn	Gravitational wave detectors and experiments
% 98.54.-h	Quasars; active or peculiar galaxies, objects, and systems
%Use showkeys class option if keyword
\maketitle

\section{Introduction}

The recent discovery of gravitational waves (GWs) from stellar-mass binary black hole mergers by the Laser Interferometer Gravitational-wave Observatory (LIGO; \cite{2016PhRvL.116f1102A}) opened the door to alternative probes of stellar and galactic evolution, cosmology and fundamental physics. As LIGO's sensitivity will gradually increase in the coming years, and other GW detectors come online \cite{TheVirgo:2014hva,PhysRevD.88.043007,India}, the rate of GW observations will increase to $\sim1$\,day$^{-1}$ \cite{PhysRevX.6.041015,2016LRR....19....1A}, making it possible to comprehensively study GW source populations.

The majority of detected GWs will come from binary black hole (BBH) mergers \cite{PhysRevX.6.041015}, making it possible to study these sources in detail. At the same time, while there is significant effort invested in finding electromagnetic or neutrino emission from GW sources \cite{2016ApJ...826L..13A,2016PhRvD..93l2010A,2013CQGra..30l3001B}, it is not clear if any BBH mergers will have detectable multimessenger signatures. Such counterparts can be produced only if the binary is surrounded by sufficient gas that can be accreted, which is not expected for most formation channels (but see \cite{2016arXiv160203831B,2017MNRAS.464..946S,2016ApJ...821L..18P,2016ApJ...822L...9M,2016ApJ...819L..21L}).

BBHs can be formed either from massive stellar binaries \citep{2002ApJ...572..407B,2012ApJ...759...52D,Marchant16,Mandel16} or via dynamical interactions in dense stellar systems, including globular clusters and galactic nuclei \citep{2002ApJ...576..899P,millerhamilton02,2004Natur.428..724P,2009MNRAS.395.2127O,2012PhRvD..85l3005K,2013ApJ...773..187N,Antonini2014ApJ...781...45A,Antognini2014MNRAS.439.1079A,2015ApJ...800....9M}. Identifying the BBH formation channel will be a key step in using BBH mergers as cosmic probes. However, this identification for a single merger is difficult without an electromagnetic counterpart \cite{PhysRevX.6.041015}. Possible observational clues include the binary mass ratio \cite{PhysRevX.6.041015}, mass distribution \cite{2016ApJ...824L..12O}, black-hole spin \cite{2015arXiv150304307V,2016ApJ...832L...2R}, and orbital eccentricity close to merger \cite{2009MNRAS.395.2127O}. However, the efficiency of these clues in differentiating between formation channels is uncertain, and is dependent on complex stellar evolution and dynamics processes.

Accurate GW localization, and the identification of the host galaxy \cite{1986Natur.323..310S}, could be an additional important clue in constraining the formation channel. However, the large number of possible host galaxies will require highly accurate localization that may only be achieved for a small fraction of GW observations \cite{2016arXiv161201471C}, or with future GW detector networks. The situation is complicated further if multiple formation channels are present, ultimately requiring the statistical study of rare events.

In this paper we investigate the prospects of statistically proving the connection of BBH mergers with rare hosts, focusing on luminous AGN. %While star formation and stellar mass are present in all galaxies--translating to a high number-density host population, BBH mergers may preferentially occur in a small subset of galaxies. For example,
Copious gas inflow to the nuclei of bright AGN (i.e. quasars) provides a potential site for finding massive black holes \cite{2014MNRAS.441..900M} that can enhance the merger rate as BBHs embedded in their accretion disks rapidly merge due to gas dynamical friction \cite{2016arXiv160203831B,2017MNRAS.464..946S}. Active galactic nuclei (AGN) represent a small fraction ($\sim 1\%$) of galaxies, making their identification as hosts feasible. We will establish the feasibility of statistically proving the connection with AGN as hosts for a set of observed BBH mergers, even if only a sub-population of the detected mergers originate from AGN.  This technique can probe the physical origin of BBH mergers, and allow cosmological measurements of the Hubble constant using plausible AGN host redshifts (as previously proposed for extreme mass ratio inspiral events in galaxies detectable by LISA; \cite{MacLeodHogan2008}) without EM counterparts.
%ZH: I added the above
The method can be extended and applied to other similarly rare host populations.

\vspace{-\baselineskip}
\section{Search strategy}

We aim to take advantage of spatial correlation between the distribution of a host population, namely AGN, and the location of origin of detected GWs. The origin of each detected GW can be localized to within a finite volume at high confidence \cite{2016ApJ...829L..15S}. GWs originating from an AGN population will preferentially come from regions in the universe with higher AGN number density, while GWs of other origin will show no such preference.

The source population is assumed to be a known set of point sources. Since Advanced LIGO-Virgo will only be sensitive to mergers at redshift $z \lesssim 1$ \cite{2010CQGra..27q3001A}, it is possible to achieve a sufficiently complete AGN catalog within this range. Additionally, even the closest AGN have much smaller angular diameter than the precision of GW reconstruction, hence we can treat AGN as point sources. We will further assume that AGN are randomly, uniformly distributed within the local universe, a conservative assumption given that they are known to cluster \cite{2009ApJ...697.1634R}. We will also assume that AGN are spatially uncorrelated with alternative merger sources (see further discussion of this point below).

A network of GW detectors can constrain the location of origin of a GW, generating a 3D probability density. This probability distribution is typically expressed as a 3D localization comoving volume (hereafter localization volume), at a threshold confidence or credible level (CL), often chosen to be 90\% \cite{2016LRR....19....1A}. For simplicity, we will use this localization volume at 90\% CL, without taking advantage of the inner probability density.

Let the localization volume of the $i^{th}$ detected GW be $V_i$. For a GW \emph{not} originating from an AGN, the number of AGN within $V_i$ will follow a Poisson distribution with mean $\lambda_i = \rho_{\rm agn}V_i$, where $\rho_{\rm agn}$ is the number density of AGN. The probability of having $N_{\rm agn,i}$ AGN's within $V_i$ for our null-hypothesis is
\begin{equation}
  \mathcal{B}_i(N_{\rm agn,i}) = \mbox{Poiss}(N_{\rm agn,i}, \rho_{\rm agn}V_i).
  \label{eq:PoissBG}
\end{equation}
If the GW originated in an AGN, then there will be 1 guaranteed AGN, with the number of additional AGN following a Poisson distribution with $\lambda_i$ mean. The probability of having $N_{\rm agn,i}$ AGN in $V_i$ volume for our alternative hypothesis is therefore
\begin{equation}
  \mathcal{S}_i(N_{\rm agn,i}) = \mbox{Poiss}(N_{\rm agn,i}-1, \rho_{\rm agn}V_i).
  \label{eq:PoissS}
\end{equation}

While a single GW detection may not be sufficient to determine its host population of origin, combining multiple GW observations will increase the likelihood that we can statistically prove the connection between BBH mergers and AGN. For the alternative hypothesis that a fraction $\mathrm{f}_{\rm agn}$ of the detected GWs originated from AGN and $(1-\mathrm{f}_{\rm agn})$ from some other host population, we obtain the likelihood
\begin{equation}
  \mathcal{L}(\mathrm{f}_{\rm agn}) = \prod_{i} \left[\mathrm{f}_{\rm agn} \mathcal{S}_i + (1-\mathrm{f}_{\rm agn})\mathcal{B}_i\right]
\end{equation}
where the product is over all detected BBH mergers during the observation period.

While we do not know $\mathrm{f}_{\rm agn}$ \emph{a priori}, here we will use a specific value in the likelihood ratio test. For estimating $\mathrm{f}_{\rm agn}$, one can maximize $\mathcal{L}(\mathrm{f}_{\rm agn})$ with respect to $\mathrm{f}_{\rm agn}$ (e.g., \cite{2008APh....29..299B}). Since we focus here on sensitivity and not the precision of parameter reconstruction, we will ignore this step.

The test statistic of a set of detected GWs will be the likelihood ratio
\begin{equation}
  \lambda = 2\log\left[\frac{\mathcal{L}(\mathrm{f}_{\rm agn})}{\mathcal{L}(0)}\right]
  \label{eq:lambda}
\end{equation}
The significance of an ensemble of GW observations will be determined by comparing the observed $\lambda$ value to $\lambda$'s background distribution $P_{\rm bg}(\lambda)$. This background distribution could in principle be determined by direct integration, but for simplicity, we used Monte Carlo simulations very similar to our previous work \cite{2016arXiv161103861B}.
%ZH: I added the above.  I am 99 percent sure this could be done without Monte Carlo, since all the distributions are known
We will reject the null hypothesis in favor of $\mathrm{f}_{\rm agn}$ fraction of the GWs originating from AGN if the GWs' $\lambda$ corresponds to a p-value less than $3\sigma$ ($=0.00135$).

We now determine how many GW detections will be sufficient to reject the null hypothesis with $3\sigma$ significance, given an $\mathrm{f}_{\rm agn}$ fraction. We characterize this number, $N_{\rm gw,3\sigma}(\mathrm{f}_{\rm agn})$, by the \emph{median} number of detections that is sufficient to reject the null hypothesis at $3\sigma$ level.

\subsection{Monte Carlo Simulations}

To find $N_{\rm gw,3\sigma}(\mathrm{f}_{\rm agn})$, we use Monte Carlo simulations (see \cite{2016arXiv161103861B}). We adopt the distribution of GW localization volumes obtained by Chen \& Holz \cite{2016arXiv161201471C} for the LIGO-Virgo GW detector network at design sensitivity. We take their results for 10\,M$_\odot$--10\,M$_\odot$ BBH mergers as a characteristic binary, noting that the localization volume increases for greater binary mass due to the lower GW frequencies as well as the larger horizon distances. The localization volumes found by Chen \& Holz are given at $90\%$ confidence level. As only $90\%$ of GWs will be located within these localizations volumes, this decreases the \emph{effective} AGN fraction $\mathrm{f}_{\rm agn}$ by a factor of 0.9, which we take into account in the following calculations.

For one Monte Carlo realization, we assume a GW detection number $N_{\rm gw}$, and (effective) AGN fraction $\mathrm{f}_{\rm agn}$. For each GW detection, we randomly assign an AGN origin with $0.9\mathrm{f}_{\rm agn}$ probability, otherwise it is considered to originate from another host type. For GW detection $i$, we assign a random localization volume $V_i$ drawn from the Chen-Holz distribution described above, and generate a random AGN number within this volume, using the distributions in Eqs. \ref{eq:PoissS} and \ref{eq:PoissBG} for AGN and not-AGN origins, respectively. We then calculate $\lambda$ corresponding to this realization using Eq. \ref{eq:lambda}. We similarly obtain background realizations following the same calculation, except that the number of AGN for all GW detections are drawn from the background distribution in Eq. \ref{eq:PoissBG}. We use these background realizations to determine the p-values corresponding to the realizations with GWs from AGN.

We repeat the above Monte Carlo analysis for a range of $N_{\rm gw}$ values for a given $\mathrm{f}_{\rm agn}$ in order to determine the threshold number $N_{\rm gw,3\sigma}(\mathrm{f}_{\rm agn})$.

\section{Results}
\label{section:results}
% money plot and description

We carried out the Monte Carlo analysis described above for a range of $\mathrm{f}_{\rm agn}$ values. We found the scaling relation
\begin{equation}\label{eq:fractionscaling}
  N_{\rm gw,3\sigma} \propto \mathrm{f}_{\rm agn}^{-2}.
\end{equation}
This can be intuitively expected: the standard deviation of the total number of AGN for all GW detections together scales with $N_{\rm gw,3\sigma}^{1/2}$, while the number of ''signal" AGN scales with $N_{\rm gw,3\sigma}\mathrm{f}_{\rm agn}$, corresponding to a signal-to-noise ratio $\mbox{SNR}\propto N_{\rm gw,3\sigma}^{1/2}\mathrm{f}_{\rm agn}$. Interpreting our detection threshold as a fixed SNR, we get back Eq. \ref{eq:fractionscaling}.

\begin{figure}
  \centering
  \includegraphics[width=0.48\textwidth]{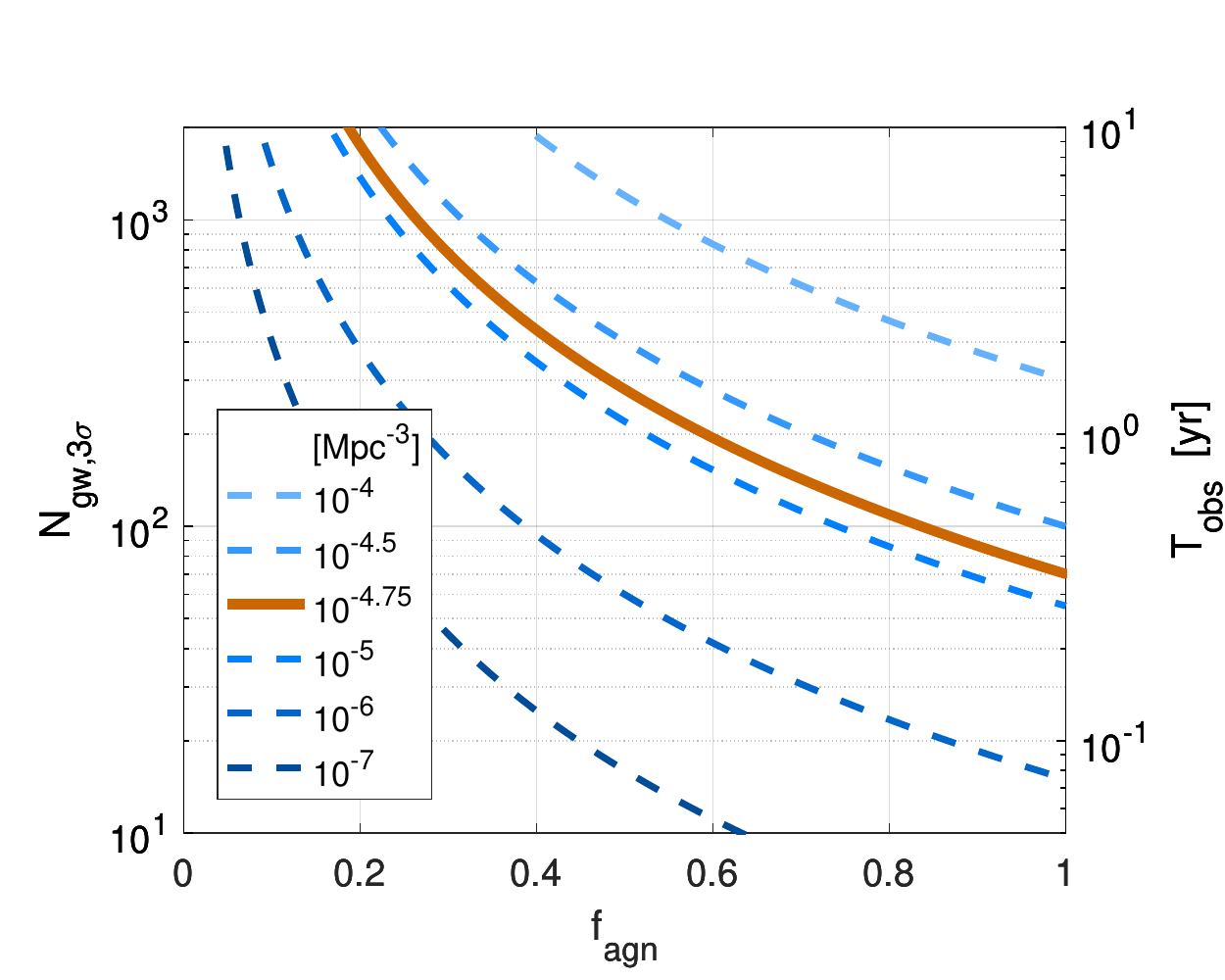}
  \caption{Number of detections needed for the identification of an AGN host population at a median $3\sigma$ significance, as a function of the fraction ($\mathrm{f}_{\rm agn}$) of GW detections originating from AGN. The results are shown for different assumed AGN number densities (see legend), with fiducial density $\rho_{\rm agn}=10^{-4.75}$\,Mpc$^{-3}$ \cite{2007ApJ...667..131G,2009ApJ...704.1743G}, for the Advanced LIGO-Virgo network at design sensitivity. On the right side, we mark the necessary observation duration corresponding to numbers of detection, using 200 detections/year. }\label{figure:detectiontime}
\end{figure}

Our results for $N_{\rm gw,3\sigma}$ are shown in Fig. \ref{figure:detectiontime}. Here, we applied Monte Carlo analysis to evaluate the case $\mathrm{f}_{\rm agn}=1$, and used Eq. \ref{eq:fractionscaling} to show scaling with $\mathrm{f}_{\rm agn}$.

For a fiducial AGN density $\rho_{\rm agn}=10^{-4.75}$\,Mpc$^{-3}$ \cite{2007ApJ...667..131G,2009ApJ...704.1743G}, we find that $\sim 70$ detections would be sufficient to statistically prove the BBH-AGN connection if all BBH mergers occurred in AGN. For the expected BBH merger rate of $\sim 60$\,Gpc$^{-3}$yr$^{-1}$ \cite{PhysRevX.6.041015}, this corresponds to a few months of LIGO-Virgo observation time at design sensitivity, indicating that this scenario would yield results quickly (possibly even before LIGO and Virgo reach design sensitivity). With the current uncertainty in BBH merger rate of $9-240$\,Gpc$^{-3}$yr$^{-1}$ \cite{PhysRevX.6.041015}, the required time is within a month and a few years. Even if only a fraction of mergers occur in AGN, we find that a 5-year observation period is likely sufficient to statistically prove the BBH-AGN connection if the AGN fraction is at least $25\%$.

To understand our results' dependence on the uncertain AGN number density, and to demonstrate the method's applicability to other rare host types, we also obtained results for a range of different $\rho_{\rm agn}$ number densities. While a sparser source population leads to even less detections needed for the identification of a host population, we found that, even for number densities of $10^{-4}$\,Mpc$^{-3}$, a 5-year observation period will establish the BBH-AGN connection for $\mathrm{f}_{\rm agn}\gtrsim0.5$. With higher host number densities it becomes difficult to prove an AGN connection solely using GW localizations, although the construction of additional GW detectors (KAGRA \cite{2013PhRvD..88d3007A} and LIGO India \cite{LIGOIndia}), and potentially the inclusion of marginally significant GW events, will further improve the situation.

How much do GWs with different localization volumes contribute to the results? To better understand the role of the GW localization volume, we carried out the Monte Carlo analysis described above, but with fixed localization volume size. This analysis confirmed that, for fixed localization volume, our method described above is equivalent to combining AGN from all GW detections, and looking for a $3\sigma$ deviation in the overall number. The GW detection number threshold in this case, for $N_{\rm gw,3\sigma} \gg 1$, is $ N_{\rm gw,3\sigma} = 9\rho_{\rm agn}V$, where $V$ is the fixed localization volume.

Are we mainly relying on a few well-localized GWs for this analysis, or are less-well-localized GWs also useful? To answer this question, we reran our Monte Carlo study for $\rho_{\rm agn}=10^{-4.75}$\,Mpc$^{-3}$ and $\mathrm{f}_{\rm agn}=1$, with the modification that the analysis only used GW localization volumes below a cutoff volume $V_{\rm cutoff}$. We measured $N_{\rm gw,3\sigma}$ as a function of the fraction $\mathrm{f}_{\rm V}(V_{\rm cutoff})$ of localization volumes below $V_{\rm cutoff}$. We found that $V_{\rm cutoff}\gtrsim 10^5$\,Mpc$^{3}$ did not meaningfully change $N_{\rm gw,3\sigma}$, but for lower $V_{\rm cutoff}$, we found a quick deterioration.  %We conclude that using the top $\sim5\%$ best localized GW sources (i.e. those 3-4 events in whose error boxes we expect $\lesssim 2$ interloper quasars on average) drives our constraints. This percentage is greater (smaller) for lower (higher) host densities.
We conclude that those events drive our constraints in whose localization volumes we expect $\lesssim 2$ interloper quasars on average. For $\rho_{\rm agn}=10^{-4.75}$\,Mpc$^{-3}$, this is the top $\sim5\%$ best localized events.

\section{Discussion and Conclusions}
% conclude that this is promising
% summarize next steps in terms of even more accurate analysis (BBH mass distribution, different detector network, AGN correlation, unequal weighting of AGN)
% quasar uniform distribution

We examined the prospects of using GW localization to probe GW host galaxy populations. In particular, we were interested in BBH mergers in AGN, where the inflow of gas can significantly increase merger rates compared to other galactic nuclei, and may lead to multimessenger emission. AGN represent a small fraction of galaxies, making it easier to identify correlation between their position and the localization of GWs. We considered the Advanced LIGO-Virgo detector network at its design sensitivity.

We calculated the number of GW observations needed to statistically establish the connection of BBH mergers with AGN as hosts, as a function of the AGN number density $\rho_{\rm agn}$ and the fraction $\mathrm{f}_{\rm agn}$ of BBH mergers originating in AGN. Our findings are the following:

\noindent \emph{(1)} For our fiducial number density $\rho_{\rm agn}=10^{-4.75}$\,Mpc$^{-3}$ and $\mathrm{f}_{\rm agn}=1$, $\sim 70$ observations will be sufficient, on average, to statistically establish the BBH-AGN connection at $3\sigma$ significance. With an expected rate of BBH merger detection of $200$\,yr$^{-1}$, this corresponds to 6 months of observation time.
%ZH: why 6 months?   200/7 = 3, so sounds like 4 months.  Is this a typo or did you include some down-time fraction for the instrument ?

\noindent \emph{(2)} We quantified the efficiency of establishing BBH-AGN for fractional AGN contributions (Fig. \ref{figure:detectiontime}). For a fractional contribution $\mathrm{f}_{\rm agn}$, the number of detections sufficient on average to establish the BBH-AGN connection is $N_{\rm gw,3\sigma} = 70\, \mathrm{f}_{\rm agn}^{-2}$.

\noindent \emph{(3)} Incorporating the $\sim5\%$ best localized BBH mergers is sufficient to produce close to optimal constraints.

Our results demonstrate that correlation between the location of AGN and GWs \emph{alone} can be used to establish the BBH-AGN connection with a few years of observation with Advanced LIGO-Virgo at design sensitivity, making this a competitive approach compared to more model dependent probes that use reconstructed BBH properties, or uncertain electromagnetic counterparts.

In principle, LIGO events can be spatially correlated with AGN even if they are unrelated to AGN, but occur in galaxies whose spatial distribution is correlated with AGN.  The cross-correlation length between local galaxies and quasars is $\sim 6$\,Mpc \cite{2013ApJ...778...98S}, which is an order of magnitude smaller than the linear size of the typical LIGO error volume \cite{2016arXiv161201471C}, so we expect this effect to be small, unless the events occur in rare galaxy sub-types that have a stronger correlation with AGN.

The technique and results presented here can also be applied to other rare host populations. For instance, in the standard dynamical formation scenario, BBHs form in globular clusters. The number of globular clusters within a galaxy strongly correlates with the mass of the central supermassive black hole \cite{2010ApJ...720..516B}, so relatively rare, large black holes near the turnover in the Schechter function are the dominant contributors. Another relevant channel is BBHs formed by GW capture in the dense stellar mass black hole populations of galactic nuclei \cite{2009MNRAS.395.2127O}. This scenario preferentially occurs in the densest nuclei, and therefore could be strongly correlated with E+A galaxies. E+As are post-starburst galaxies that represent $\approx0.2\%$ of all $z\approx 0$ galaxies but host an order unity fraction of stellar tidal disruption events \cite{2014ApJ...793...38A, 2016ApJ...818L..21F}; preliminary evidence suggests that this is due to overdense central star clusters \cite{2016ApJ...825L..14S}.

There are important next steps that will further enhance the prospects of our analysis. (i) AGN are relatively strongly clustered, which can enhance the signal to noise ratio over a random AGN distribution assumed here. (ii) Different host populations can be spatially correlated, which needs to be taken into account before the BBHs can be inferred to reside in AGN (rather than just correlated with them statistically).
%when we aim to  distinguish between them.
%ZH: I changed the above. Is this what you meant?
(iii) Combining our error region analysis with other ''observables" (binary mass, spins, etc.) and host galaxy properties that correlate with binary rate will significantly enhance search sensitivity (iv) If suitable models are available, reconstructed BBH properties can add to the differentiating power of a search.

We further emphasize the need for detailed AGN catalogs out to redshift of $z\sim 0.2$; these catalogs will provide the backbone of any statistical BBH-AGN connection \cite{2008PhRvD..77d3512M,2015ApJ...801L...1B}. In principle, the nearby AGN which host most LIGO events, with $M_{\rm bh}\gtrsim 10^6 {\rm M_\odot}$) should be bright enough to be detectable in large all-sky surveys \cite{2009ApJ...700.1952H}.
%For reference, a $3\times10^5 {\rm M_\odot}$ black hole radiating at only 1\% of the Eddington luminosity at redshift $z=0.1$, assuming a 10\% bolometric correction, corresponds to an optical magnitude of $i\approx19$\,mag \cite{2009ApJ...700.1952H}, roughly the completeness threshold of the SDSS survey.
In practice, however, existing spectroscopically confirmed quasar catalogs appear incomplete.  In the full comoving volume out to $z=0.1$ ($\sim 0.3 {\rm Gpc^3}$) we expect to find $\sim 10,000$\,AGN.  This number is based on extrapolating spectroscopic measurements for faint nearby AGN to the rest of the sky.  On the other hand, the latest quasar catalog from SDSS, covering $\approx 10,000\,{\rm deg^2}$ \cite{2016arXiv160806483P} contains only 232 quasars at $z<0.1$, implying that it is only 10\% complete. Deeper spectroscopic surveys exist but target only a small fraction of the sky. The Swift BAT 70-month survey covers the entire sky, but contains only a total of 523 quasars and AGN at $z\leq 0.1$, the vast majority of which are Seyfert galaxies \cite{2013ApJS..207...19B}. More targeted cataloging efforts are needed to take full advantage of GW observations. With no full-sky catalogs, it is also sufficient to identify galaxies in follow-up surveys of individual GW localization volumes \cite{2015ApJ...801L...1B,Metzger+13}.

We thank Jules Halpern for useful discussions. IB, ZM and SM are thankful for the generous support of Columbia University in the City of New York. ZH acknowledges support from NASA grant NNX15AB19G and a Simons Fellowship for Theoretical Physics.  BDM acknowledges support from NASA grant NNX16AB30G and the Research Corporation for Science Advancement Scialog Program grant RCSA23810. NCS acknowledges support by NASA through Einstein Postdoctoral Fellowship Award Number PF5-160145.

%\bibliographystyle{h-physrev}
%\bibliography{Refs_BBH_AGN}

\end{document}